\PassOptionsToPackage{unicode}{hyperref}
\PassOptionsToPackage{hyphens}{url}
\documentclass[
]{article}
\usepackage{amsmath,amssymb}
\usepackage{iftex}
\ifPDFTeX
  \usepackage[T1]{fontenc}
  \usepackage[utf8]{inputenc}
  \usepackage{textcomp} 
\else 
  \usepackage{unicode-math} 
  \defaultfontfeatures{Scale=MatchLowercase}
  \defaultfontfeatures[\rmfamily]{Ligatures=TeX,Scale=1}
\fi
\usepackage{lmodern}
\ifPDFTeX\else
\fi
\IfFileExists{upquote.sty}{\usepackage{upquote}}{}
\IfFileExists{microtype.sty}{
  \usepackage[]{microtype}
  \UseMicrotypeSet[protrusion]{basicmath} 
}{}
\makeatletter
\@ifundefined{KOMAClassName}{
  \IfFileExists{parskip.sty}{%
    \usepackage{parskip}
  }{
    \setlength{\parindent}{0pt}
    \setlength{\parskip}{6pt plus 2pt minus 1pt}}
}{
  \KOMAoptions{parskip=half}}
\makeatother
\usepackage{xcolor}
\usepackage{graphicx}
\makeatletter
\def\maxwidth{\ifdim\Gin@nat@width>\linewidth\linewidth\else\Gin@nat@width\fi}
\def\maxheight{\ifdim\Gin@nat@height>\textheight\textheight\else\Gin@nat@height\fi}
\makeatother
\setkeys{Gin}{width=\maxwidth,height=\maxheight,keepaspectratio}
\makeatletter
\def\fps@figure{htbp}
\makeatother
\providecommand{\tightlist}{%
  \setlength{\itemsep}{0pt}\setlength{\parskip}{0pt}}
\ifLuaTeX
  \usepackage{selnolig}  
\fi
\usepackage[]{natbib}
\nocite{*}
\usepackage{bookmark}
\IfFileExists{xurl.sty}{\usepackage{xurl}}{} 
\hypersetup{
  pdftitle={Training on Data Analysis Reproducibility via Containerization with Apptainer},
  hidelinks,
  pdfcreator={LaTeX via pandoc}}

\title{Training on Data Analysis Reproducibility via Containerization
with Apptainer}
\author{}
\date{\today}

\begin{document}
\maketitle

\author{
  \href{https://orcid.org/0000-0002-7205-0790}{Roy Cruz Candelaria}\textsuperscript{1},
  \href{https://orcid.org/0000-0003-4033-6716}{Wouter Deconinck}\textsuperscript{2},
  \href{https://orcid.org/0000-0003-2631-9696}{Aman Desai}\textsuperscript{3},
  \href{https://orcid.org/0000-0001-8605-9772}{Guillermo Fidalgo Rodríguez}\textsuperscript{4},
  \href{https://orcid.org/0000-0002-6322-5587}{Michel Hernandez Villanueva}\textsuperscript{5},
  \href{https://orcid.org/0000-0003-2792-7511}{Kilian Lieret}\textsuperscript{6},
  \href{https://orcid.org/0000-0002-0630-5185}{Valeriia Lukashenko}\textsuperscript{7},
  \href{https://orcid.org/0000-0002-6356-2655}{Sudhir Malik}\textsuperscript{8},
  \href{https://orcid.org/0000-0002-9489-2681}{Marco Mambelli}\textsuperscript{9},
  \href{https://orcid.org/0009-0001-8158-3644}{Tetiana Mazurets}\textsuperscript{8},
  \href{https://orcid.org/0000-0001-8415-2543}{Alexander Moreno Briceño}\textsuperscript{10},
  \href{https://orcid.org/0000-0003-3282-2634}{Andres Rios-Tascon}\textsuperscript{6},
  \href{https://orcid.org/0000-0002-4656-4683}{Richa Sharma}\textsuperscript{8}
  \\[1.5em] 
  \small\textsuperscript{1}University of Wisconsin-Madison
  \small\textsuperscript{2}University of Manitoba 
  \small\textsuperscript{3}Adelaide University 
  \small\textsuperscript{4}University of Alabama
  \small\textsuperscript{5}Brookhaven National Laboratory 
  \small\textsuperscript{6}Princeton University 
  \small\textsuperscript{7}University of Zurich
  \small\textsuperscript{8}University of Puerto Rico Mayaguez 
  \small\textsuperscript{9}Fermi National Accelerator Laboratory 
  \small\textsuperscript{10}Universidad Antonio Nariño
}

\section{Summary}\label{summary}

We present the material and resources developed for training physicists
on containerization technologies enabled by Apptainer. In the context of
analysis preservation using Apptainer's capabilities, we have developed
examples that execute common tools in High Energy Physics (HEP) and
Nuclear Physics within containers. Training physicists on
containerization technologies is of utmost importance in today's
research landscape. By embracing these technologies, users can achieve
enhanced reproducibility, portability, collaboration, and resource
efficiency, assuring the conditions and integrity of the scientific
analysis process. This training module,
\href{https://hsf-training.github.io/hsf-training-singularity-webpage/}{``Introduction
to Apptainer/Singularity''}, is part of the HEP Software Foundation
Training Center, which aims to equip newcomers to the field of High
Energy Physics with the necessary software skills and best practices.

\section{Statement of Need}\label{statement-of-need}

The high complexity of the HEP analyses creates major challenges in
terms of capturing and preserving the analysis and the knowledge
surrounding it. In addition, the high value of the data produced by the
experiments demands having a framework in place for data and knowledge
preservation that allows reuse, reinterpretation, and reproducibility of
research outcomes. Containerization plays a crucial role in modern
software development and scientific research. Containers encapsulate the
entire environment, including dependencies, libraries, and
configurations, ensuring that the software or research environment can
be replicated accurately across different systems. This reproducibility
eliminates the infamous ``works on my machine'' problem and enables
others to reproduce results and build upon existing work easily.
Therefore, concepts such as reproducibility, preservation, and
distribution are enabled by containerization.

Within the wide range of solutions available, Apptainer, formerly known
as Singularity \citep{kurtzer2017singularity}, is a container platform
designed by and for scientists. It runs across various Linux operating
systems and computing environments without special requirements or
superuser permissions, simplifying deployment across different cloud
platforms, clusters, and High-Performance Computing (HPC) systems,
thereby enabling efficient use of computing resources and reducing the
time and effort required to scale analysis workflows. Apptainer has been
designed to use single-file-based container images, facilitating
distribution, archiving, and sharing. The containers can run as a
regular application, simplifying the integration with resource managers
and distributed computing environments. Additionally, the containers
preserve the permissions in the environment: the user outside the
container can be the same user inside, preventing any security concerns
from system administrators.

By training collaborators on Apptainer, experiments can establish a
common platform and language for sharing and collaborating on analysis
projects. Containers serve as self-contained units that can be easily
shared, enabling the reproducibility of results across teams,
institutions, and even globally when providing open data sets accessible
to everyone. The preservation of the analysis is ensured for years to
come, independently of the availability of dependencies for the
analysis, as long as Apptainer is supported in future operating system
versions. Therefore, it is of the utmost importance to train individuals
in containerization technologies due to the numerous benefits they
offer, such as enhanced reproducibility and seamless portability.

\section{Curriculum}\label{curriculum}

\subsection{Learning objectives}\label{learning-objectives}

The training module covers basic concepts of containerization on
Apptainer. Students learn about images and containers, how they differ
from virtual machines, and the design goals behind Apptainer to develop
reproducible environments that can be executed on different platforms in
the context of analysis preservation.

Specific goals for this training module include how to:

\begin{itemize}
\tightlist
\item
  Pull Apptainer images from libraries
\item
  Run commands inside Apptainer containers
\item
  Build Apptainer containers with user requirements and from a single
  file definition
\item
  Share files from the host system to the Apptainer container and vice
  versa
\end{itemize}

\subsection{Prerequisites}\label{prerequisites}

Students are required to have a basic knowledge of the Unix Shell
commands. In addition, they must have either access to a computing
system with Apptainer available, such as an institutional cluster, or
install Apptainer locally on a Linux system, on a Mac using Lima and
Qemu, or in WSL on Windows machines. The HEP Software Foundation (HSF),
a global community that facilitates collaboration and common efforts in
the development and sustainability of software for high-energy physics
\citep{HEPSoftwareFoundation:2017ggl}, provides in its
\href{https://hsf-training.org/training-center/}{training center} the
basic material to cover the prerequisites listed above. The Setup
section in the training module covers Apptainer's installation on the
different platforms.

\subsection{Contents}\label{contents}

The training module (or lesson) is divided into episodes (individual
chapters). Every episode has stated learning objectives, a main body
that includes exercises with solutions, and a summary in key points to
ensure the accomplishment of the learning objective. These are the
episodes with their corresponding specific objectives:

\begin{itemize}
\tightlist
\item
  \textbf{Introduction}: Learning the containerization concepts and
  design goals behind Apptainer.
\item
  \textbf{Containers and Images}: Learning to search and pull images
  from the Sylab Singularity library and Docker Hub, and interacting
  with the containers using the command line interface.
\item
  \textbf{Building Containers}: Downloading, assembling, and modifying
  containers from available images in the repositories.
\item
  \textbf{Containers from Definition Files}: Creating a container from a
  definition file.
\item
  \textbf{Sharing Files between Host and Container}: Mapping directories
  on your host system to directories within your container and learning
  about the bind paths included automatically in all containers.
\item
  \textbf{Apptainer Instances}: Running containers in a detached mode to
  keep services up and deploying instances via definition files.
\item
  \textbf{Bonus Episode: Building and Deploying an Apptainer Container
  to GitHub Packages}: Using GitHub actions to build an Apptainer
  container and share it via GitHub Container Registry (GHCR).
\end{itemize}

The examples and exercises use ROOT \citep{Antcheva:2009zz} and Python,
which are widely used in the HEP and Nuclear Physics communities.
Students practice the usage of containers to run ROOT interactively
through PyROOT \citep{Galli:2020boj}, building containers with the
Pythia8 physics event generator \citep{Bierlich:2022pfr} and the Uproot
library \citep{Pivarski_Uproot} for reading ROOT files, and creating
definition files that execute RooFit tutorials \citep{Verkerke:2003ir}.
In later episodes, participants deploy long-running services such as a
Jupyter notebook server with ROOT support and learn to share data
between the host and the container using bind mounts.

The training module is designed to be used as an aide by instructors
teaching live or to be used by students learning independently. To
facilitate asynchronous learning, each episode also includes a recording
where the instructor explains the material and guides students through
the exercises, similar to a live class.

\section{Teaching Experience}\label{teaching-experience}

This module on Apptainer has been used during the HSF \& IRIS-HEP
\citep{IRIS-HEP-webpage} training events on
\href{https://indico.cern.ch/event/1508102/}{Analysis Reproducibility}
(formerly known as \href{https://indico.cern.ch/event/1219810/}{Analysis
Preservation} and
\href{https://indico.cern.ch/event/1375507/\%5D}{Analysis Pipelines}).
During a week, we covered tools that help the participants to integrate
containerization into their scientific workflows for enhanced
reproducibility. Our focus was to guide researchers with strong
backgrounds in data analysis through the practical aspects of using
Apptainer to encapsulate their analysis pipeline. Participants learned
the basic concepts of containerization and how they can be applied in
the context of data-intensive workflows.

Through pre-recorded lectures and hands-on exercises, participants
experience firsthand how to build and run Apptainer containers, convert
existing Docker images into Apptainer for HPC environments, and manage
software dependencies for long-term analysis preservation, all with
resources they are familiar with. During the live mentoring sessions and
support via Slack, we helped participants troubleshoot setup issues,
review exercises, and understand how they can apply these tools to their
own analysis. This interactive component was essential for tailoring our
advice to the computing environment available to the participants.

To collect metrics related to the overall learning experience, we have
implemented a two-step survey process, circulated among participants
before and after the training events. The pre-event feedback enables
educators to tailor their approach to meet the unique needs of each
group, ensuring that attendees receive explanations at a proper level of
complexity. The post-event surveys assess the effectiveness of the
training event and gather valuable information on what worked well and
what parts require revision. This feedback loop is fundamental for the
continuous improvement of the material and the training events. All
registered participants are encouraged to complete the whole survey,
including the questions about sections that they did not attend or
complete. To ensure a sufficient response rate to the surveys while
maintaining anonymity, we implemented an anonymous verification process:
submitting the survey generates a return code that can be entered into
the Indico system to confirm that the survey was completed.

Figure 1 shows the pre-survey data collected during the registration for
the training events from 360 registered participants between 2023 and
2025, providing insights into user familiarity with various
Singularity/Apptainer commands and concepts. The significant majority of
respondents had ``Never heard of it'' for most of the listed Apptainer
commands and topics. This trend is consistent across all categories,
indicating that most participants are researchers with no exposure to
Apptainer, strongly motivating the need for training. Figure 2
illustrates the post-survey data collected from 82 participants after
the training events. The plots show a visible increase in the knowledge
of the students when compared with the pre-survey. Users generally
reported being ``I am familiar with'' or ``Very familiar'' with core
commands like apptainer pull, apptainer shell, and apptainer exec. This
indicates a good grasp of fundamental operations. There's a significant
number of users who selected ``Never heard of it'' for certain commands,
particularly '' apptainer instance''. This suggests the core learning
objectives are met, while `apptainer instance' serves as an advanced
topic. Future iterations could explore participant interest in this
feature more directly. On the other hand, it is important to remind that
we encourage all event attendees, including those who did not start or
complete this training module, to fill out the complete survey. This may
explain the people who answered ``Never heard of it'' or ``Used it
once''.

\begin{figure}
\centering
\includegraphics{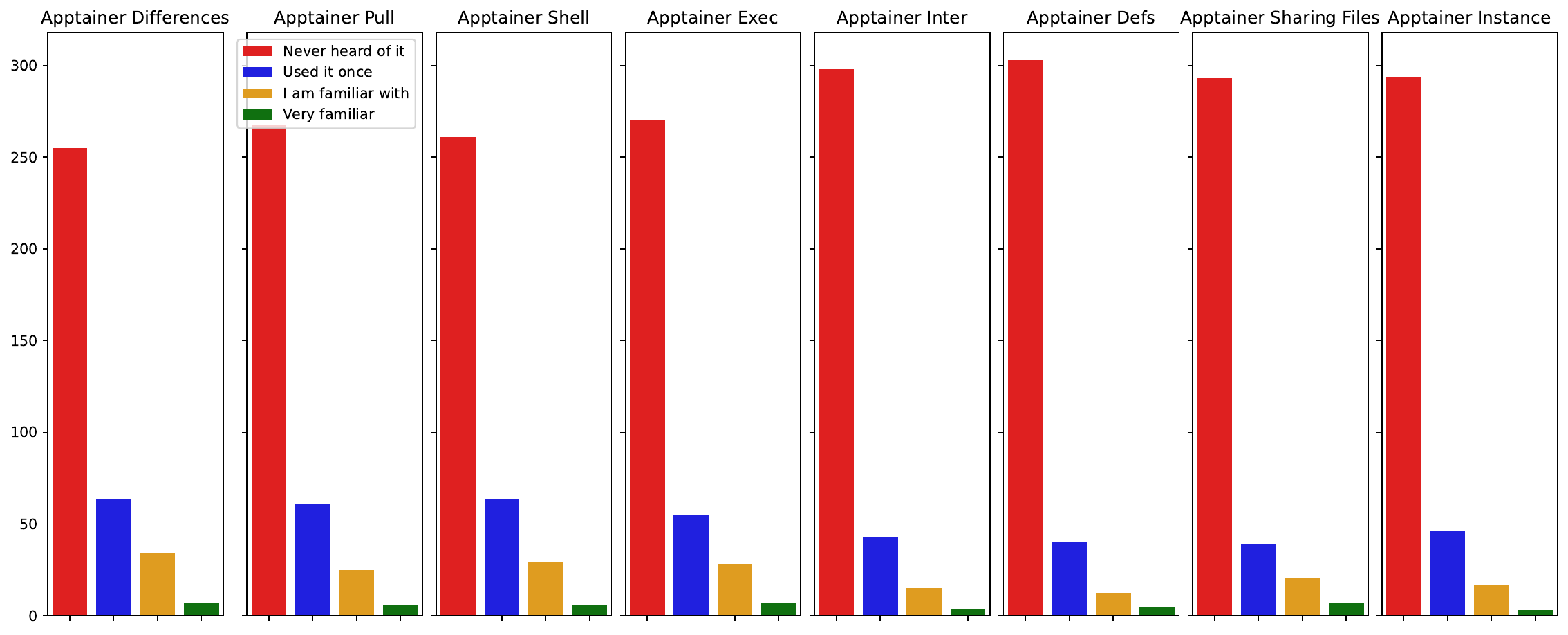}
\caption{Pre-survey data collected during the training event
registration, asking about familiarity with various Apptainer commands
and concepts \label{Figure_1}}
\end{figure}

\begin{figure}
\centering
\includegraphics{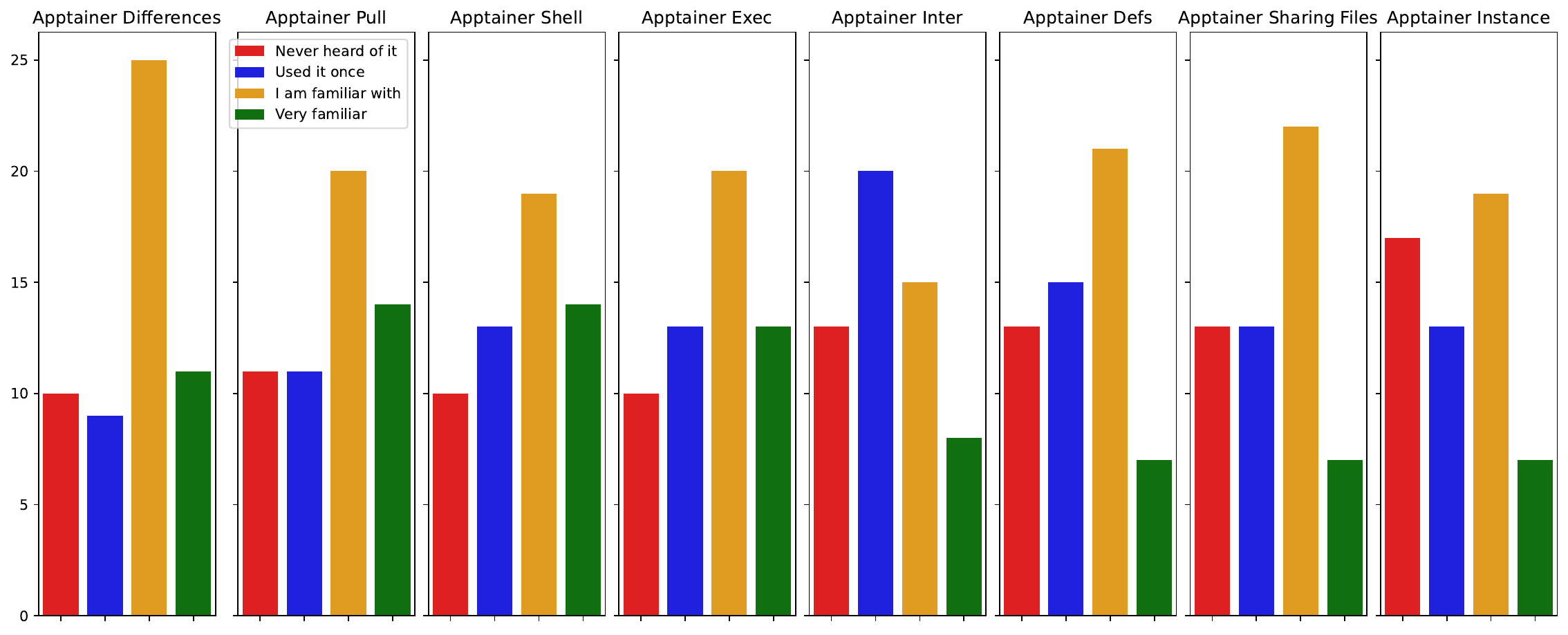}
\caption{Post-survey data collected, asking about familiarity with
various Apptainer commands and concepts after the training event
\label{Figure_2}}
\end{figure}

Trainees largely agreed that the material had a proper difficulty level
and enough exercises to feel an interactive experience, as shown in the
Figure 3. Similarly, most respondents felt that the number of exercises
was ``About right'' for interactive learning. This suggests that the
balance of theoretical content and practical application was effective.

\begin{figure}
\centering
\includegraphics{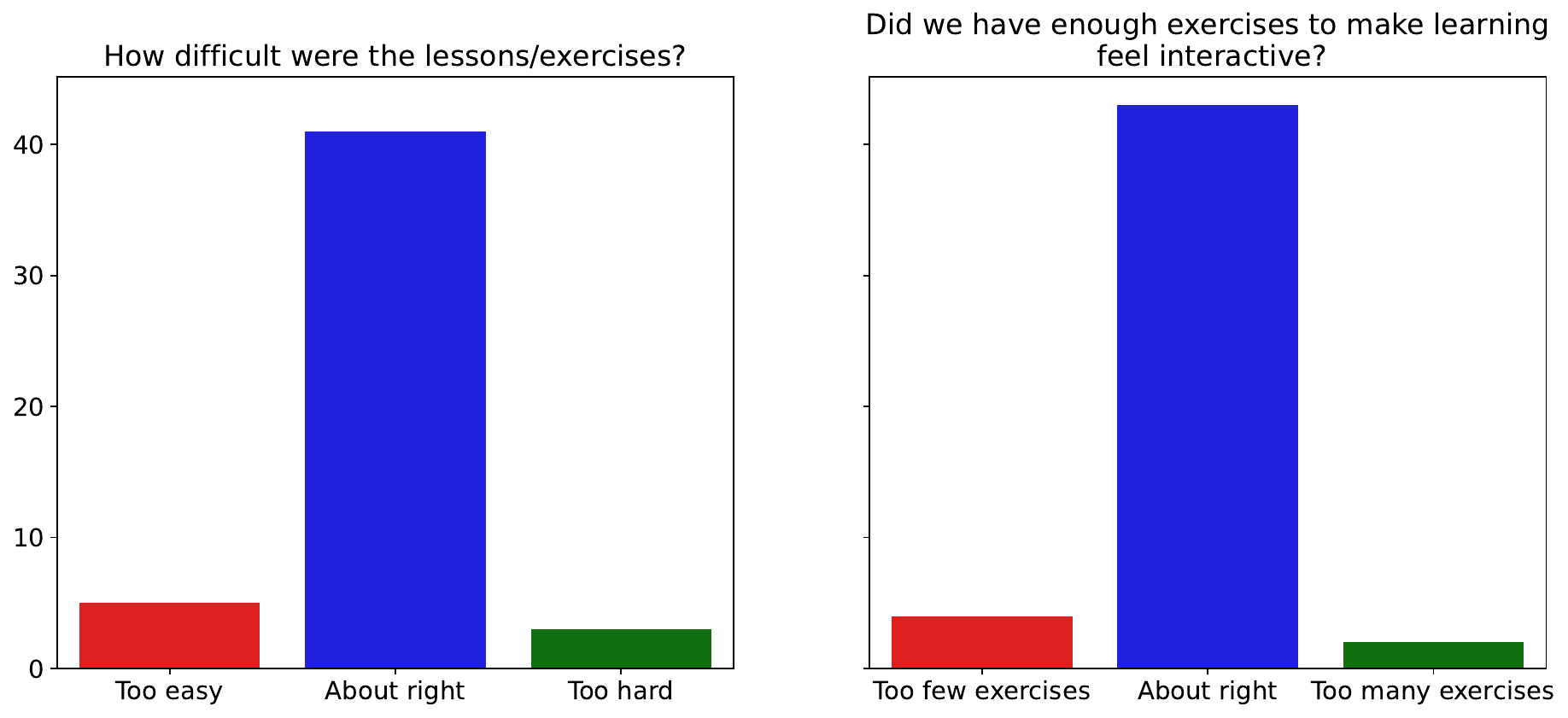}
\caption{Metrics from the pre-event (post-event) surveys
\label{Figure_3}}
\end{figure}

\section{Conclusions}\label{conclusions}

This training module successfully addresses the need to introduce a
reproducible, portable, and resource-efficient data analysis in High
Energy and Nuclear physics using Apptainer. The module equips students
and researchers with the skills to create and manage a reproducible
analysis environment, covering a range of concepts from the fundamentals
of containerization to the use of containers, the creation of images,
and the best practices to prepare and share a reproducible analysis.

The positive feedback consistently observed in the pre- and post-event
surveys demonstrates a significant increase in the knowledge of
participants and satisfaction with the material's difficulty and
exercises. The demonstrated success of this module in various training
events, reaching 360 registered participants, illustrates its
contribution to fostering a platform for scientific collaboration and
ensuring the long-term preservation of complex analyses, independent of
evolving environments.

\section{Acknowledgements}\label{acknowledgements}

We would like to thank all the members of our community in the HEP
Software Foundation and IRIS-HEP training for their voluntary
contributions, big or small. We thank NSF grants PHY-2323298 ,
OAC-1836650, OAC-1829707, and OAC-1829729 for support of the training
programs. This work was supported by the U.S. Department of Energy under
contract number DE-SC0012704.

\renewcommand\refname{References}
  \bibliography{paper.bib}

@article{kurtzer2017singularity,
    doi = {10.1371/journal.pone.0177459},
    author = {Kurtzer, Gregory M. AND Sochat, Vanessa AND Bauer, Michael W.},
    journal = {PLOS ONE},
    publisher = {Public Library of Science},
    title = {Singularity: Scientific containers for mobility of compute},
    year = {2017},
    month = {05},
    volume = {12},
    url = {https://doi.org/10.1371/journal.pone.0177459},
    pages = {1-20},
    number = {5},
}

@article{HEPSoftwareFoundation:2017ggl,
    author = "Albrecht, Johannes and others",
    collaboration = "HEP Software Foundation",
    title = "{A Roadmap for HEP Software and Computing R{\&}D for the 2020s}",
    eprint = "1712.06982",
    archivePrefix = "arXiv",
    primaryClass = "physics.comp-ph",
    reportNumber = "HSF-CWP-2017-01, HSF-CWP-2017-001, FERMILAB-PUB-17-607-CD",
    doi = "10.1007/s41781-018-0018-8",
    journal = "Comput. Softw. Big Sci.",
    volume = "3",
    number = "1",
    pages = "7",
    year = "2019"
}

@article{Antcheva:2009zz,
    author = "Antcheva, I. and others",
    title = "{ROOT: A C++ framework for petabyte data storage, statistical analysis and visualization}",
    eprint = "1508.07749",
    archivePrefix = "arXiv",
    primaryClass = "physics.data-an",
    reportNumber = "FERMILAB-PUB-09-661-CD",
    doi = "10.1016/j.cpc.2009.08.005",
    journal = "Comput. Phys. Commun.",
    volume = "180",
    pages = "2499--2512",
year = "2009"
}

@article{Galli:2020boj,
    author = "Galli, Massimiliano and Tejedor, Enric and Wunsch, Stefan",
    editor = "Doglioni, C. and Kim, D. and Stewart, G. A. and Silvestris, L. and Jackson, P. and Kamleh, W.",
    title = "{A New PyROOT: Modern, Interoperable and More Pythonic}",
    doi = "10.1051/epjconf/202024506004",
    journal = "EPJ Web Conf.",
    volume = "245",
    pages = "06004",
    year = "2020"
}

@article{Verkerke:2003ir,
    author = "Verkerke, Wouter and Kirkby, David P.",
    editor = "Lyons, L. and Karagoz, Muge",
    title = "{The RooFit toolkit for data modeling}",
    eprint = "physics/0306116",
    archivePrefix = "arXiv",
    reportNumber = "CHEP-2003-MOLT007",
    journal = "eConf",
    volume = "C0303241",
    pages = "MOLT007",
    year = "2003"
}

@article{Bierlich:2022pfr,
    author = "Bierlich, Christian and others",
    title = "{A comprehensive guide to the physics and usage of PYTHIA 8.3}",
    eprint = "2203.11601",
    archivePrefix = "arXiv",
    primaryClass = "hep-ph",
    reportNumber = "LU-TP 22-16, MCNET-22-04, FERMILAB-PUB-22-227-SCD",
    doi = "10.21468/SciPostPhysCodeb.8",
    journal = "SciPost Phys. Codeb.",
    volume = "2022",
    pages = "8",
    year = "2022"
}

@software{Pivarski_Uproot,
    author = "Pivarski, Jim and others",
    title = "{Uproot}",
    url = "https://github.com/scikit-hep/uproot5",
    doi = "10.5281/zenodo.4340632",
    year = "2017"
}

@misc{IRIS-HEP-webpage,
    author = "IRIS-HEP",
    title = "{Institute for Research and Innovation in Software for High Energy Physics}",
    howpublished = "\url{http://iris-hep.org}"
}

\end{document}